# GMap: Drawing Graphs as Maps

Emden R. Gansner, Yifan Hu, and Stephen G. Kobourov

AT&T Research Labs, Florham Park, NJ, {erg,yifanhu,skobourov}@research.att.com

**Abstract.** Information visualization is essential in making sense out of large data sets. Often, high-dimensional data are visualized as a collection of points in 2-dimensional space through dimensionality reduction techniques. However, these traditional methods often do not capture well the underlying structural information, clustering, and neighborhoods. In this paper, we describe GMap: a practical tool for visualizing relational data with geographic-like maps. We illustrate the effectiveness of this approach with examples from several domains All the maps referenced in this paper can be found in `www.research.att.com/~yifanhu/GMap`.

## 1 Introduction

Graphs capture relationships between objects and graph drawing allows us to visualize such relationships. Typically vertices are placed as points in two or three dimensional space, and edges are represented as lines between the corresponding vertices. The layout optimizes some aesthetic criteria, for example, minimal edge length and edge crossings. While such point-and-line representation are most commonly used, other representations have also been considered. For example, treemaps [27] use a recursive space filling approach to represent trees. There is also a large body of work on representing planar graphs as *contact graphs* [7, 12, 19], where vertices are embodied by geometrical objects and edges are shown by two objects touching in some specified fashion. Koebe's theorem [16] shows that all planar graphs can be represented by touching disks. Similar representation is possible with triangles, where two adjacent vertices correspond to vertex-to-side touching pair of triangles, as shown by de Fraysseix *et al.* [7]. If vertices are represented by rectilinear regions and edges correspond to side-to-side contact between paired regions, He [12] has shown that all planar graphs have such drawings. Graph representations of side-to-side touching regions tend to be visually appealing and have the added advantage that they suggest the familiar metaphor of a geographical map.

In this paper we describe GMap, an algorithm to represent general graphs as maps. Clearly, there are theoretical limitations to what graphs can be represented exactly by touching polygons, even when allowing for non-convexity. However, our aim here is practical rather than graph theoretical. We do not insist that the created map be an exact representation of the graph but that it captures the underlying relationships well. With this in mind, we do not insist that all vertices are represented by individual polygons either. In fact, we group closely connected vertices into regions. If we would like to show all of the relationships, we can superimpose a graph drawing on top of the map.

Our overall goal is to create a representation which makes the underlying data understandable and visually appealing. Our map representation is especially effective when the underlying graph contains structural information such as clusters and/or hierarchy. The traditional line-and-point representation of graphs often requires considerable effort to comprehend, and often puts off general users. On the other hand, a map representation is more intuitive, as most people are very familiar with maps and even enjoy carefully examining maps.

Given that we do not insist on the map to be an exact representation for the graph, at first it may seem trivial to generate a map. For example, one can start with a “good” graph drawing and build a Voronoi diagram of the vertices along with the four corners of the bounding box for the drawing. However, the results are visually unappealing as maps, with straight borders between "countries" and jagged, angular overall appearance. Our GMap algorithm takes as input a graph and produces a map with a "natural" look, outer boundaries that follow the outline of the vertex sets, and inner boundaries having the twists and turns found in real maps. Our maps also can have lakes, islands, and peninsulas, similar to those found in real geographic maps; see Figures 3-5.[1]

## 2 Related Work

There is little previous work on generating map representations of graphs. Most related work deals with accurately and appealingly representing a given geographic region, or on re-drawing an existing map subject to additional constraints. Examples of the first kind of problem are found in traditional cartography, e.g., the 1569 Mercator projection of the sphere into 2D Euclidean space. Examples of the second kind of problem are found in cartograms, where the goal is to redraw a map so that the country areas are proportional to some metric, an idea which dates back to 1934 [25] and is still popular today (e.g., the New York Times red-blue maps of the US, showing the presidential election results in 2000 and 2004 with states drawn proportional to population).

The map of science [4] uses vertex coloring in a graph drawing to provide an overview of the scientific landscape, based on citations of journal articles. Treemaps [27], squarified treemaps [6] and the more recent newsmaps [30] represent hierarchical information by means of space-filling tilings, allocating area proportional to some important metric.

Representing imagined places on a map as if they were real countries also has a long history, e.g., the 1930's Map of Middle Earth by Tolkien [29] and the Bücherlandes map by Woelfle from the same period [1]. More recent popular maps include xkcd's Map of Online Communities [2]. While most such maps are generated in an ad hoc manner and are not strictly based on underlying data, they are often visually appealing.

Generating synthetic geography has a large literature, connected to its use in computer games and movies. Most of the work (e.g., [20, 22]) relies on variations of a fractal model. Although these techniques could provide additional photorealism, it is unclear how they could be used with the position and size constraints attached to the maps we consider here.

## 3 The Mapping Algorithm

The input to our algorithm is a relational data set from which we extract a graph $G = (V, E)$. The set of vertices $V$ corresponds to the objects in the data, e.g., authors in the graph drawing community, and the set of edges $E$ corresponds to the relationship between pairs of objects, e.g., co-authoring a paper. In its full generality, the graph is vertex-weighted and edge-weighted, with vertex weights corresponding to some notion of the importance of a vertex and edge weights corresponding to some notion of the distance between a pair of vertices.

The first step in our GMap algorithm is to embed the graph in the plane. Possible embedding algorithms include principal component analysis [15], multidimensional scaling (MDS) [17], force-directed algorithm [10], or non-linear dimensionality reduction such as LLE [26] and Isomap [28].

[1] This paper contains zoom-able high resolution images; all the images are also available at `www.research.att.com/~yifanhu/GMap`.

The second step is a cluster analysis of the underlying graph or the embedded pointset from step one. In this step, it is important to match the clustering algorithm to the embedding algorithm. For example, a geometric clustering algorithm such as $k$-means [21] may be suitable for an embedding derived from MDS, as the latter tends to place similar vertices in the same geometric region with good separation between clusters. On the other hand, with an embedding derived from a force-directed algorithm [10], a modularity based clustering [23] could be a better fit. The two algorithms are strongly related, as pointed out in the recent findings by Noack *et al.* [24], and therefore we can expect vertices that are in the same cluster to also be geometrically close to each other in the embedding.

In the third step the two-dimensional embedding together with the clustering are used to create the map. Using the embedding information, a Voronoi diagram of the vertices is created. A naive approach would be to form the Voronoi diagram of the vertices, together with fours points on the four corners of the bounding box; see Fig. 1(a). This would result in aesthetically unappealing maps with unnatural outer boundaries and sharp corners. A more natural appearance can be obtained by placing some random points. A random point is only accepted, if its distance from any of the real points is more than $r$ (a preset threshold) away. This leads to more rounded boundaries. The randomness of the points on the outskirts also gives rise to some randomness of the outer boundaries, thus making them more realistic and natural; see Fig. 1(b). Furthermore, depending on the value of $r$, this step can also result in the creation of lakes (e.g., Fig. 5) in areas where vertices are far apart from each other. Nevertheless, some inner boundaries remain artificially straight.

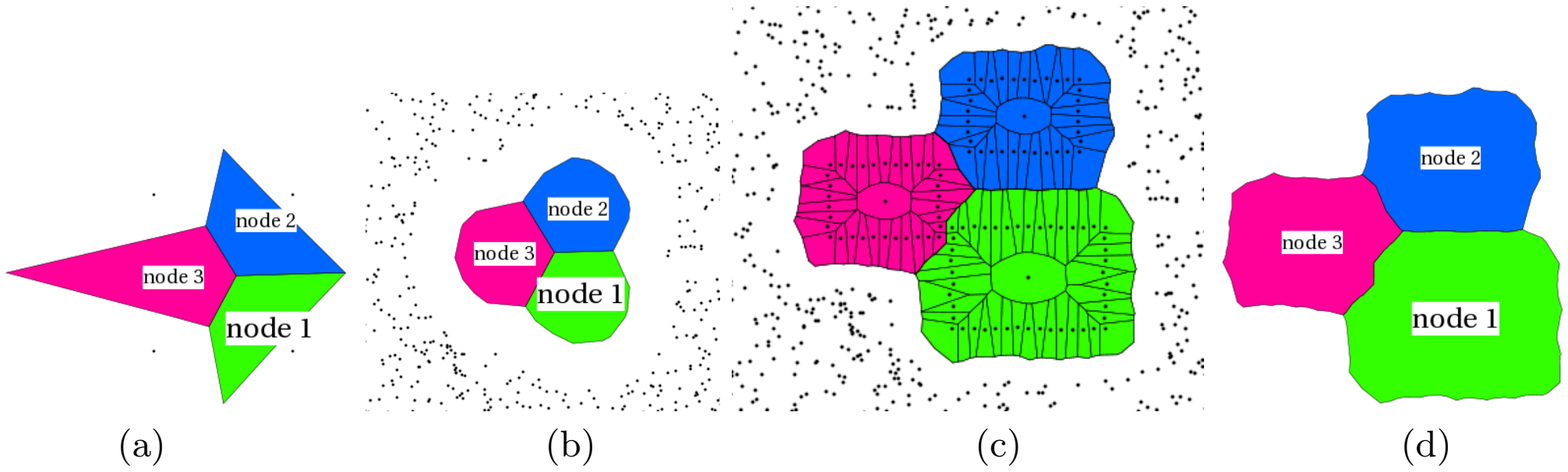

**Fig. 1.** (a) Voronoi diagram of vertices and corners of bounding box; (b) better construction of outer boundaries through placement of random points; (c) Voronoi diagram of vertices and points inserted around the bounding boxes of the labels; (d) the final map.

Another undesirable feature is that the three "countries" all have more or less equal area, whereas we might often want some areas to be larger than the others, perhaps due to importance of the entities them represent. As an illustration, in Fig. 1, we assume that the area corresponding to "node 1" is more important than the other two areas, and use a larger label for that area. To make areas proportional to the label size, we first generate artificial points along the bounding boxes of the labels; see Fig. 1(c). To make the inner boundaries more realistic, we perturb these points randomly instead of running strictly along the rectangle bounding boxes. Here Voronoi cells that belong to the same vertex are colored in the same color, and cells that correspond to the random points on the outskirt are not shown. Cells of the same color are then merged to give the final map in Fig. 1(d). Note that instead of the bounding boxes of labels, we could use any 2D shapes, e.g., the outlines of real countries, in order to obtain a desired look and proportion of area, as long as these shapes do not overlap.

When mapping vertices that contain cluster information, in addition to merging cells that belong to the same vertex, we also merge cells that belong to the same cluster, thus forming regions of complicated shapes, with multiple vertices and labels in each region. At this point we can add more geographic components to strengthen the map metaphor. For instance, in places where there is significant space between vertices in neighboring clusters, we can add lakes, rivers, or mountain ranges to the map to indicate the distance. These structures can all be formed by similar insertion of random points in places where vertices are far away from each other.

In terms of complexity, the algorithm is scalable and has a time complexity of $O(|V|\log|V|)$. We first add $n_a$ artificial points along the bounding boxes of the labels, typically $n_a = 40|V|$. We then insert $n_r$ random points of distance $r$ away from any vertices and artificial points, usually $n_r$ is set to between $|V|$ to $40|V|$, depending on the size of the graph. This step is carried out by first forming a quadtree of the vertices and artificial points, which takes time $O(|V|\log|V|)$, then testing whether a random point is within distance $r$ of the set of vertices and artificial points. Each test takes $O(\log|V|)$ time, thus overall $O(|V|\log|V|)$. We first compute a Delaunay triangulation of the points, which can be done in time $O(|V|\log|V|)$ [9]. Then we create the corresponding Voronoi diagram of all points and merge Voronoi cells that belongs to the same cluster. This step requires $O(|V|)$ and thus the overall complexity of GMap is $O(|V|\log|V|)$, with a relatively large coefficient due to the large number of artificial and random points. As a reference point, all maps in this paper were generated in a few seconds. Mapping a larger graph with $|V| = 440,000$, $n_r = |V|$ and $n_a = 40|V|$ took 4 minutes[2].

## 4 GMap Maps

In this section we examine several maps produced by our algorithm. The underlying data comes from different domains and the corresponding graphs are structurally different and of varying sizes.

### 4.1 Collaboration Graph

This graph has authors as vertices and collaborations as edges. That is, there is an edge between two authors if they have collaborated on a paper. The graph has 509 vertices and 1517 edges. The largest component has 275 vertices and 784 edges, and thus contains about 54% of all authors. The data comes from the first 10 years of the Symposium on Graph Drawing, 1994-2004. We look at the first eight largest connected components. This graph is cumulative, in the sense that two authors are connected with an edge if they have written at least one joint paper in the first ten years of the symposium. Even when drawn with a high-quality scalable force-directed algorithm [13] and after applying a node-overlap removal step, the resulting graph looks more like a hairball than anything else; see Fig. 2.

On the other hand, the corresponding map, as shown in Fig. 3, seems much more "readable". The map shows one continent corresponding to the largest connected component and seven islands, corresponding to the seven largest remaining connected components. The continent contains about a dozen "countries" determined by the collaboration patterns. The size of each label is determined by the logarithm of the number of publications and the edge thickness is similarly proportional to the number of collaborations. However, node weights and edge weights are not used in the layout calculations.

[2] Using a single thread on a Linux machine with 16 Intel Xeon processors, each with 4 cores running at 2.4 GHz, with 16 GB memory per processor.

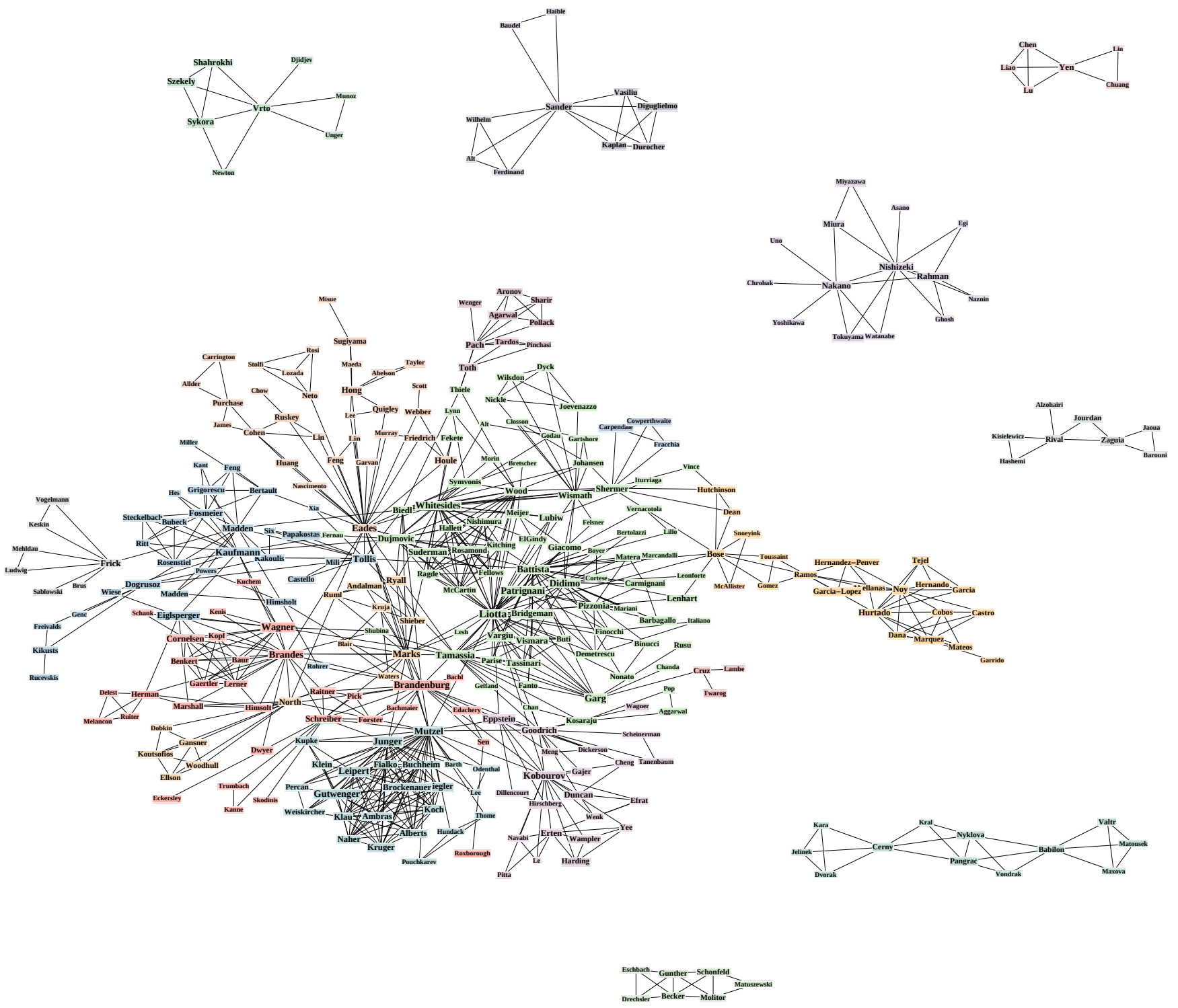

**Fig. 2.** Author collaboration graph for the GD conference, 1994-2004.

It is easy to see that European authors dominate the main continent. Several well-defined German groups can be seen on the west and southwest coasts. A largely Italian cluster occupies the center, with an adjacent Spanish peninsula in the east. The northwest contains a mostly Australasian cluster. Two North American clusters are to be found in the southeast and in the southwest, the latter one made up of three distinct components. A combinatorial geometry cluster forms the northernmost point of the main continent. Most Canadian researchers can be found in the central Italian cluster and the Spanish peninsula. Northeast of the mainland lies a large Japanese island and southeast of the mainland there is a large Czech island. Northwest of the mainland is a Crossings Number island.

### 4.2 TradeLand

Fig. 4 is a map visualizing the trade relations between all countries. Bilateral trade data between each of the 209 countries and its top trading partners were acquired from Mathematica's `CountryData` package. The font size of a label is proportional to the logarithm of the total trade volume of the country, and the color of a label reflects whether a country has a trade surplus (black) or deficit (red).

The label color gives an easy way to spot the oil-rich countries with large surpluses, which are distributed all over the world as well as in our map: Middle East (Saudi Arabia, Kuwait), Europe (Russia), South America (Venezuela), Africa (Nigeria, Equatorial Guinea). On the other hand, the countries with huge deficits are mostly in Africa (Sierra Leone, Senegal, Ethiopia) with the United States, the clear outlier.

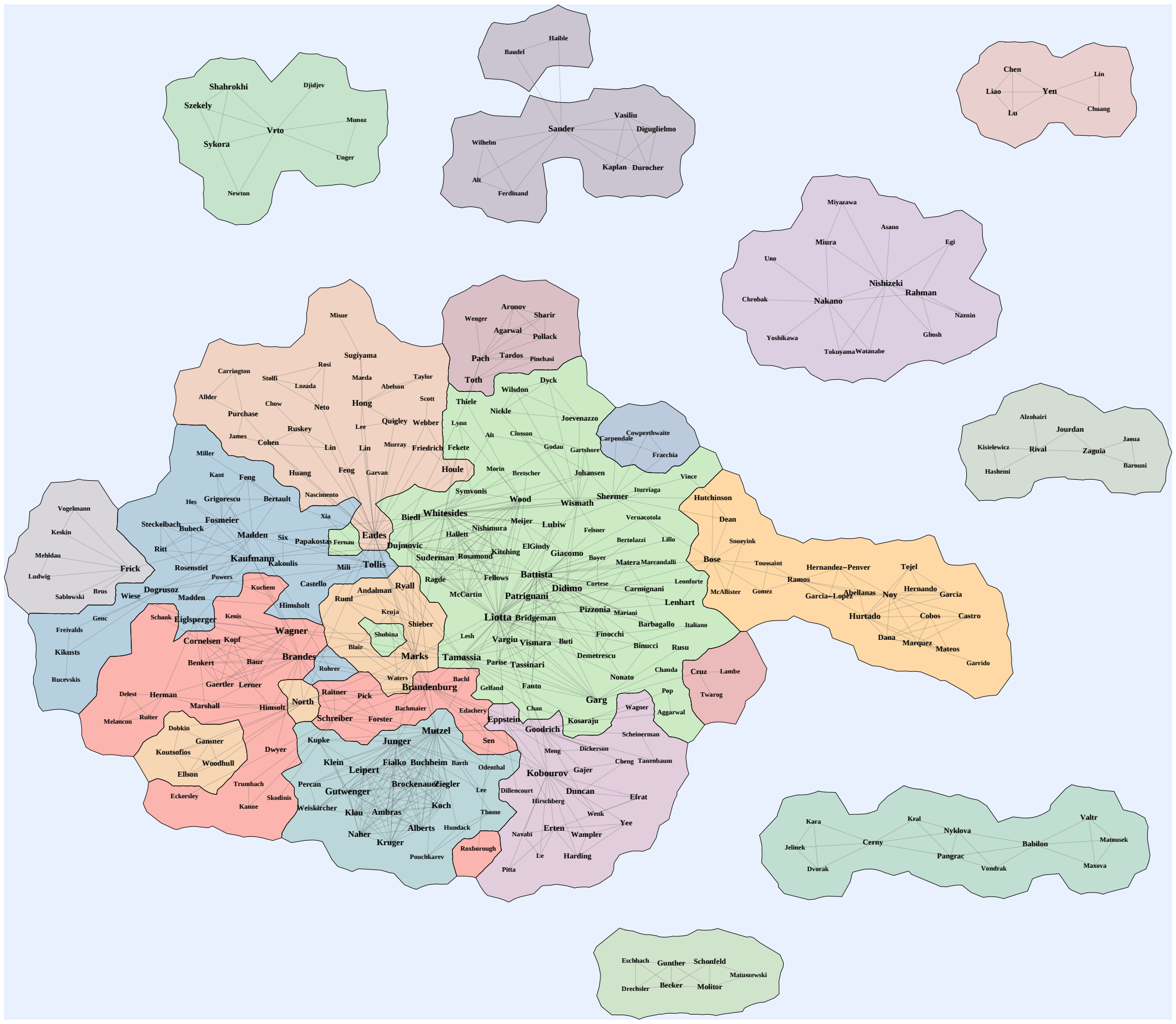

**Fig. 3.** Author collaboration map for the GD conference, 1994-2004.

Many countries in close geographic proximity end up close in our map, e.g, Central American countries like Honduras, El Salvador, Nicaragua, Guatemala and Costa Rica are close to each other in the northeast. Similarly the three Baltic republics, Latvia, Lithuania and Estonia, are close to each other in the northwest. This is easily explained by noting that geographically close countries tend to trade with each other. There are easy to spot exceptions: North Korea is not near South Korea, Israel is not particularly close to Jordan or Syria.

The G8 countries (Canada, France, Germany, Italy, Japan, Russia, United Kingdom, and the United States) are all in close proximity to each other in the center of the map. Two of the largest and closest countries in our map are China and the United States. Clearly, the proximity is due to the very large trade volume rather than geographic closeness. All these countries are in the largest cluster which is dominated by European countries in the west, Asian countries in the east, and Middle Eastern countries in the south. African countries are distributed in several clusters in close proximity to China (a major trading partner to many African countries), the United States (trading less with Africa these days), and around former colonizers (e.g., Togo, Cameroon and Senegal, which are all close to France). South American and Central American countries form several clusters in the north of the map. On the periphery of the map are small countries from around the world, and countries with few trading partners.

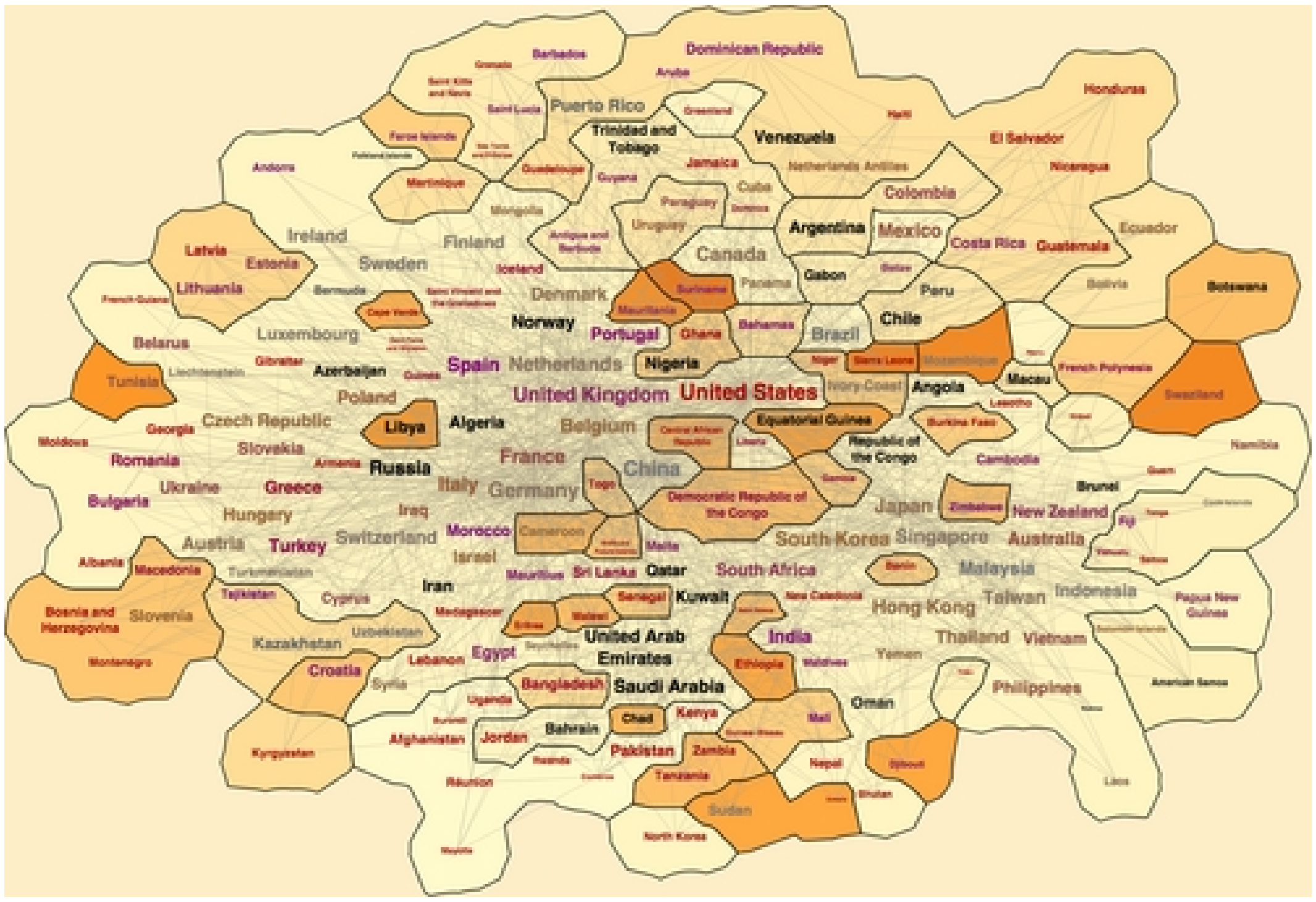

**Fig. 4.** A map of trade relations between countries.

## 4.3 BookLand Maps

Many e-commerce websites provide recommendations to allow for exploration of related items. Traditionally this is done in the form of a flat list. For example, Amazon typically lists around 5-6 books under "Customers Who Bought This Item Also Bought", with a clickable arrow to allow a customer to see further related items.

Instead of a flat list, which provides a very limited view of the neighborhood, there have been attempts to convey the underlining connectivity of the products through graph visualization. For example, *TouchGraph*, a New Jersey-based company, has an Amazon browser (`http://www.touchgraph.com/TGAmazonBrowser.html`) which essentially lays out a graph taken from a small neighborhood surrounding the book of concern. None of the existing approaches, however, gives a comprehensive view of the relationship and the clustering structures.

Using our GMap algorithm, we obtained the map in Fig. 5. The underlying data is obtained with a breadth-first traversal following Amazon's "Customers Who Bought This Item Also Bought" links, starting from the George Orwell's *1984*. All books in the map are at most 9 hops away from the source node. We further merge nodes that represent the same book, but with different publishers or different bindings. This reduces the number of vertices by 1-4%. As can be seen by the 5 versions of Chinua Achebe's *Things Fall Apart* in the central cluster, we are not always successful. The underlying graph for this map contains 913 vertices and 3410 edges. With an average degree of nearly eight, peripheral vertices in this map have only a handful of edges while central vertices have more than 20 immediate neighbors. We next examine several of the "countries" in the map in more detail. More countries are examined in the Appendix, where we also show several close-ups from the map.

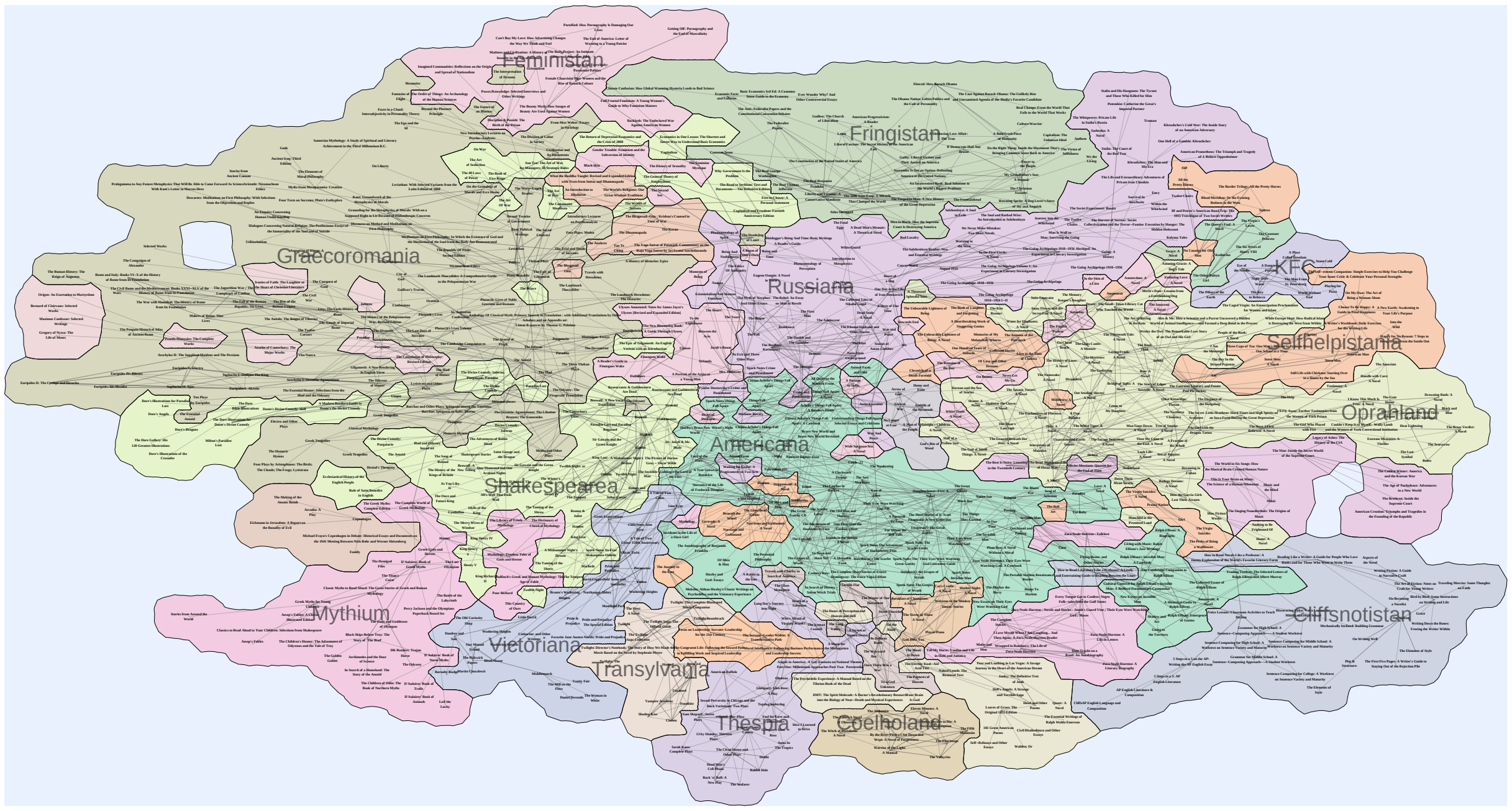

**Fig. 5.** A map of books related to "1984"

**Americana:** Somewhat surprisingly, George Orwell's *1984* along with *Animal Farm* ended up in the west corner of a region populated mostly by American writers. Britain is also represented by William Golding's *The Lord of the Flies* and Aldous Huxley's *Brave New World* along with Anthony Burgess's *Clockwork Orange*, which connect the British corner of the region to the main part dominated by 20th century American classics. Ray Bradbury's *Fahrenheit 451* and Salinger's *Catcher in the Rye* provide a transition to a variety of well-known novels: Steinbeck's *Grapes of Wrath* and *Of Mice and Men*, Ernest Hemingway's *For Whom the Bell Tolls* and *The Old Man and the Sea*, F. Scot Fitzgerald's *Great Gatsby*, Harper Lee's *To Kill a Mockingbird* and Ralph Ellison's *Invisible Man*, Joseph Heller's *Catch 22*, Kurt Vonnegut's *Slaughterhouse Five*, Ken Kesey's *One Flew Over a Cuckoo's Nest*. Some 19th century novels can also be found here: Nathaniel Hawthorne's *Scarlet Letter* and Mark Twain's *Adventures of Huckleberry Finn*.

**Victoriana:** To the southwest of Americana is a region dominated by Dickens, Austen and Bronte novels. Starting with *A Tale of Two Cities*, *Great Expectations* and *Oliver Twist* in the north and going through *Jane Eyre*, *Pride and Prejudice*, *Sense and Sensibility* and *Wuthering Heights* in the middle, the region ends with more Dickens' books in the southwest (*The Pickwick Papers*) and George Elliot novels in the southeast (*Middlemarch*).

**Russiana:** To the north of Americana lies one of the largest countries in BookLand, dominated by Russian literature and history. The core contains classic novels by Dostoyevsky (*Crime and Punishment, The Brothers Karamazov*), Tolstoy (*War and Peace, Anna Karenina*), and Solzhenitsyn (*The Gulag Archipelago, Cancer Ward*). In the northern part of the region is a collection of books about Russia and Russian history: *Stalin: The Court of the Red Tsar, Khrushchev: The Man and His Era* and *Potemkin: Catherine the Great's Imperial*

*Partner*. In the west there is a cluster of Albert Camus books (*The Stranger, The Plague, The Fall*), all well connected with the Russian classics.

**Graecoromania:** Another large region to the west of Americana contains a diverse collection of Graeco-Roman books. History books by Thucydides, Plutarch, Livy, Suetonius, Salust share the region with philosophy by St. Augustine, Plato, Socrates, and Aristotle. Greek theater is represented by Aristophanes, Aeschylus, Euripides, Sophocles and epic poetry by Homer and Virgil.

**Mythium:** Close to Graecoromania, on the southwest coast, lies the the legendary land of Mythium. *Aesop's Fables, Greek Myths for Young Children* and *The Gods and Goddesses of Olympus* are next door to *D'Aulaires' Book of Trolls, D'Aulaires' Book of Animals* and *D'Aulaires' Book of Norse Myths.*

**Shakespearea:** Very centrally located, neighboring Victoriana, Americana, Russiana, Graecoromania, and Mythium lies the land of Shakespeare. It is not surprising that nearly all tragedies, comedies and histories are present but it is interesting to observe what non-Shakespeare books are in this region: Chaucer's *Canterbury Tales*, Tennyson's *Idyls of the King*, Dante's *Divine Comedy*, *One Thousand and One Arabian Nights, Beowulf* and *The Adventures of Robin Hood.*

## 5 The Map Coloring Algorithm

In this section we consider the problem of assigning good colors to the countries in our maps. The Four Color Theorem states that only four colors are needed to color any map so that no neighboring countries share the same color. It is implicitly assumed that each country forms a contiguous region. However, this result is of limited use to us because countries in our maps are often not contiguous. For instance, a group of North American researchers are placed in a cluster made from three disjoint regions in light orange color in the southwest corner of the main continent; see Fig. 3. In cases where one cluster is represented by several disjoint regions we must use the same color for all regions to avoid ambiguity. Thus, four colors (or even five or six) are not enough.

In GMap we start with a coloring scheme from ColorBrewer [5], which typically has 5 easy to differentiate base colors, and generate as many as the number of countries by blending the base colors. As a result our color space is linear and discrete. Because of the blending, any two consecutive colors in the linear array of colors are similar to each other. When applying these colors to the map, we want to avoid coloring neighboring countries with such pairs of colors. With this in mind, we define the *country graph*, $G_c = \{V_c, E_c\}$, to be the undirected graph where countries are vertices, and two countries are connected by an edge if they share a non-trivial boundary. We then consider the problem of assigning colors to nodes of $G_c$ so that the color distance between nodes that share an edge is maximized.

More formally, let $C$ be the color space, i.e., a set of colors; let $c : V_c \to C$ be a function that assigns a color to every vertex; and let $w_{ij} \geq 0$ be weights associated with edges $\{i, j\} \in E_c$. Let $d : C \times C \to R$ be a color distance function. Define the vector of color distances along edges to be

$$v(c) = \{w_{i,j}\ d(c(i), c(j)) \mid \{i, j\} \in E_c\}.$$

Then we are looking for a color function that maximizes this vector with respect to some cost function. Two natural cost functions are:

$$\max_{c \in C}\{\sum_{\{i,j\} \in E_c} w_{i,j}\ d(c(i), c(j))^2\} \quad \text{(2-norm), or} \quad \max_{c \in C}\{\min_{\{i,j\} \in E_c} w_{i,j}\ d(c(i), c(j))\} \quad \text{(MaxMin)}$$

Dillencourt et al. [8] investigated the case where all colors in the color spectrum are available. They proposed a force directed model aimed at selecting $|V_c|$ colors as far apart as possible in the color space. However in our map coloring problem, for aesthetic reasons, we are limited to "map-like" colors, and our color space is discrete. Therefore we model our coloring problem as one of vertex labeling, where our color space is $C = \{1, 2, \ldots, |V_c|\}$, and the color function we are looking for is a permutation that maximizes the labeling differences along the edges. The cost functions we consider are

$$\max \sum_{\{i,j\} \in E_c} w_{i,j}(c_i - c_j)^2, \; c \text{ is a permutation of } \{1, 2, \ldots, |V_c|\} \quad \text{(2-norm)} \tag{1}$$

and

$$\max \min_{\{i,j\} \in E_c} w_{i,j}|c_i - c_j|, \; c \text{ is a permutation of } \{1, 2, \ldots, |V_c|\} \quad \text{(MaxMin)}$$

where $c_i$ is the $i$-th element of the vector $c$.

The complementary problem of finding a permutation that *minimizes* the labeling differences along the edges is well-studied. For example, in the context of minimum bandwidth or wavefront reduction ordering for sparse matrices, it is known that the problem is NP-hard, and a number of heuristics [14,18],were proposed. One such heuristic is to order vertices using the Fiedler vector. Motivated by this approach, we approximate (1) by

$$\max \sum_{\{i,j\} \in E_c} w_{i,j}(c_i - c_j)^2, \text{ subject to } \sum_{k \in V_c} c_k \; = \; 1 \tag{2}$$

where $c \in R^{|V_c|}$. This continuous problem is solved when $c$ is the eigenvector corresponding to the largest eigenvalue of the weighted Laplacian of the country graph, while the Fiedler vector (the eigenvector corresponding to the second smallest eigenvalue) minimizes the objective function above. Once (2) is solved, we use the ordering of the eigenvector as an approximate solution for (1). We call this algorithm SPECTRAL.

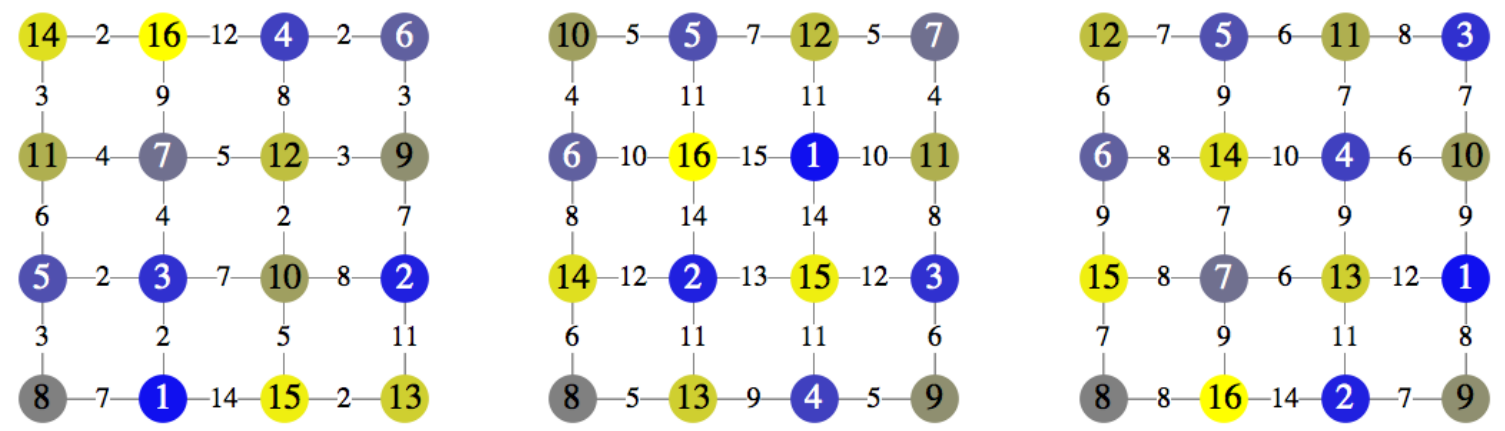

**Fig. 6.** Coloring schemes RANDOM, SPECTRAL, and SPECTRAL+GREEDY. Each node is colored by the color index shown as the node label. Edge labels are the absolute difference of the endpoint labels.

Fig. 6 illustrates three coloring schemes on a 4-4 unweighted grid graph given 16 colors in the Blue-Yellow spectrum. A random assignment of colors, RANDOM, does reasonably well, but has one edge with a color difference of 2. SPECTRAL performs better, with the minimum color difference of 4. However there are still 2 edges with a color difference of only 4. It is easy to see that SPECTRAL can be improved (e.g., swapping colors 6 and 2 would improve the measurements according to both cost functions). With this in mind we developed GREEDY, a greedy refinement algorithm based on repeatedly swapping pairs of vertices, provided that the swap improves the coloring scheme according to one of the two cost functions. Starting from a coloring scheme obtained by SPECTRAL and applying GREEDY often leads to significant improvements.

The GREEDY algorithm has a high computational complexity as we consider all possible $O(|V_c|^2)$ pairs of vertices for potential swapping. Since recomputing the cost functions can be done in time proportional to the sum of degrees of the pair on nodes considered for swapping, the overall complexity of GREEDY is $O(|V_c|^2 + |E_c|^2)$. Because the country graph $G_C$ is typically much smaller than the underlying graph $G$, GREEDY is still quite fast and all maps in this paper were colored using SPECTRAL+GREEDY.

## 6 Conclusion and Future Work

In this paper we described GMap, an efficient algorithm for drawing graphs as geographic maps. Using a number of structurally different graphs and graphs of different sizes, we illustrated the aesthetic appeal of the map metaphor for displaying underlying structures and clustering information. While the approach of visualizing relational information with the aid of geographical maps is general, here we showed one particular implementation where a scalable force-directed layout algorithm was coupled with a modularity-based clustering algorithm. Exploring different combinations of layout and clustering algorithms is one clear direction for future work.

While our algorithm is efficient and can handle large graphs, the resulting maps look best on large wall-sized posters and display walls. To make such maps more useful for interactive exploration of large underlying data sets we plan to incorporate topological clustering which would allow us to show the map in varying level of detail. We can leverage previous work on large graph visualization such as topological fisheye views [11] and the related compound fisheye views [3].

We plan to explore the map coloring problem further through the use of weighted graphs to promote color differences not only between neighboring countries, but also non-neighboring countries that are geographically close. In addition, the algorithm in [8] may be adapted for this problem by using a 1D color space; at the same time it would be interesting to use the spectral algorithm with three largest eigenvectors as an approximate solution for the continuous color assignment problem in 3D as studied in [8].

There are practical and theoretical obstacles to obtaining "perfect" maps, that is, maps that do not omit or distort the underlying information. However, a similar drawback plagues any 2-dimensional representation of data that is not 2-dimensional, including the standard geographical maps of Earth. Clearly, in dense graphs it is impossible to realize all graph adjacencies as neighboring countries. For example, with 8 countries we can have at most 18 pairwise neighbors (from Euler's formula for planar graphs), possibly forming some unavoidable "false negative associations". It is easier to deal with "false positive associations". Such an association between two countries can be formed if they are physically adjacent in the map but there is no strong relationship between the objects in the two countries One way to alleviate such a problem is to add "rivers" or "fords" along such borders near the coasts and "mountain ranges" inland, to convey that the two sides are close but not strongly connected.

### Acknowledgments

We would like to thank Stephen North for helpful discussions. We thank Michael Jünger for the 1994-2008 GD author collaboration data used in Fig. 11.

# 7 Appendix

In this section we provide more maps obtained with our algorithm, as well as close-up images from some of the earlier maps.

## 7.1 BookLand cont.

Figures 7-8 show close-ups of some of the large countries of Americana, Victoriana, Russiana, Graecoromania, Mythium, and Shakespearea discussed earlier. Fig. 9 shows close-ups of some of the other interesting countries in BookLand, some of which we consider in more detail below.

**Transylvania:** As with all books in this map, there is a fairly short path from *1984* to *Twilight*, the teenage favorite vampire series by Stephenie Meyers. In this case, the path goes through Victoriana via *Pride and Prejudice* and *Wuthering Heights*. The other main cluster in this region is made of novels in the *House of Night* series by Cast and Cast (*Hunted, Betrayed, Marked, Chosen*, etc).

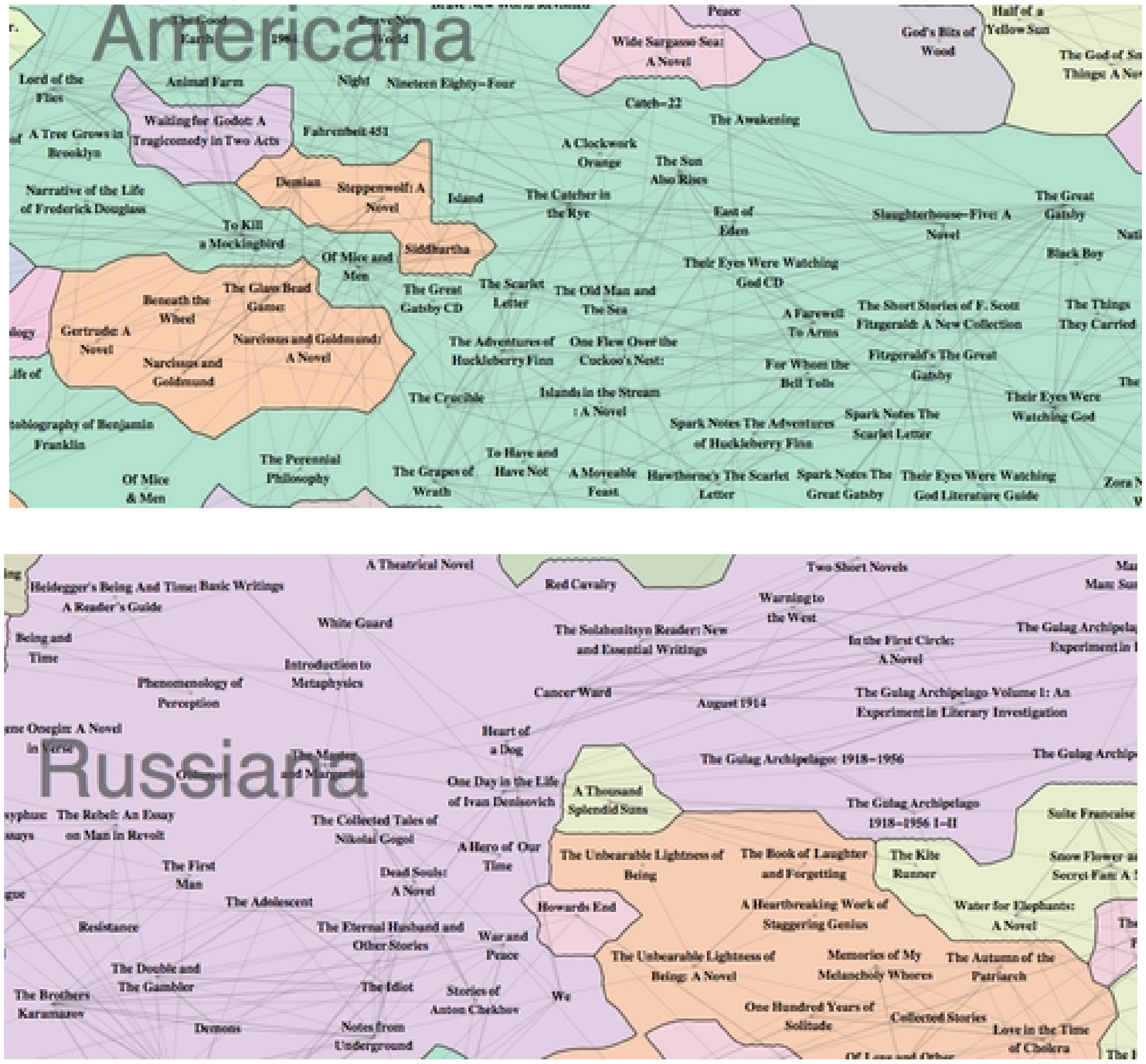

**Fig. 7.** Close-up images of various countries in BookLand'

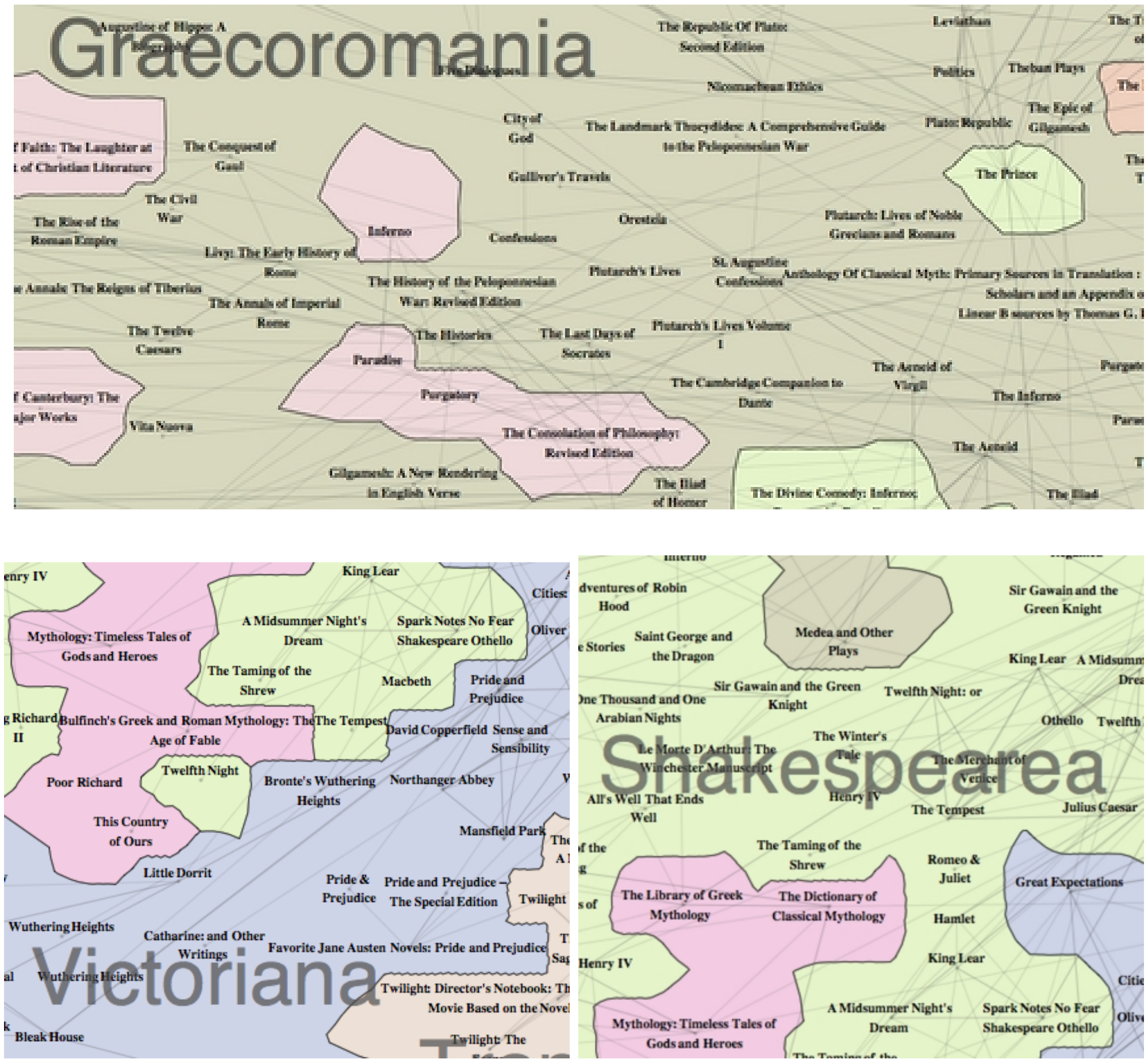

**Fig. 8.** Close-up images of various countries in BookLand'

**Thespia:** In the south, adjacent to Transylvania but mainly connected to Americana, sits a region nearly exclusively containing American Plays, from 20th century mainstays such as Tennessee Williams' *Streetcar Named Desire* and Arthur Miller's *Death of a Salesman*, through the more modern plays by David Mammet like *American Buffalo* and *Glengarry Glen Ross*, to the 2008 Broadway hit *August: Osage County*.

**Coelholand:** The very popular Brazilian writer Paulo Coelho occupies another southern region with his bestsellers *The Alchemist, Brida,* etc.

**Cliffsnotistan:** The southeast coast of BookLand contains a collection of books about writing. Classics like Strunk's *Elements of Style* and Zinnser's *On Writing Well* share the region with books such as *Sentence Composing for High School: A Worktext on Sentence Variety and Maturity.* Several Cliff's Notes books such as *5 Steps to a 5 on the AP* and *Cliff's AP English Language and Composition*, give this region its name.

**Oprahland:** A large cluster of mainly 21st century popular literature contains several "club selections" of Oprah's Book Club: *She's Come Undone, Drowning Ruth, Black and Blue.* Recent bestsellers in this region include *The Brief Wondrous Life of Oscar Wao, White Tiger, The Guernsey Literary and Potato Peel Pie Society.* Connection with Americana is through the Salman Rushdie's *Satanic Verses* and Arundhati Roy's *The God of Small Things.*

**Selfhelpistania:** An odd region, nearly contained in Oprahland, has a focus on self-help with books like *Choose to Be Happy* and *The Self-Esteem Companion: Simple Exercises to*

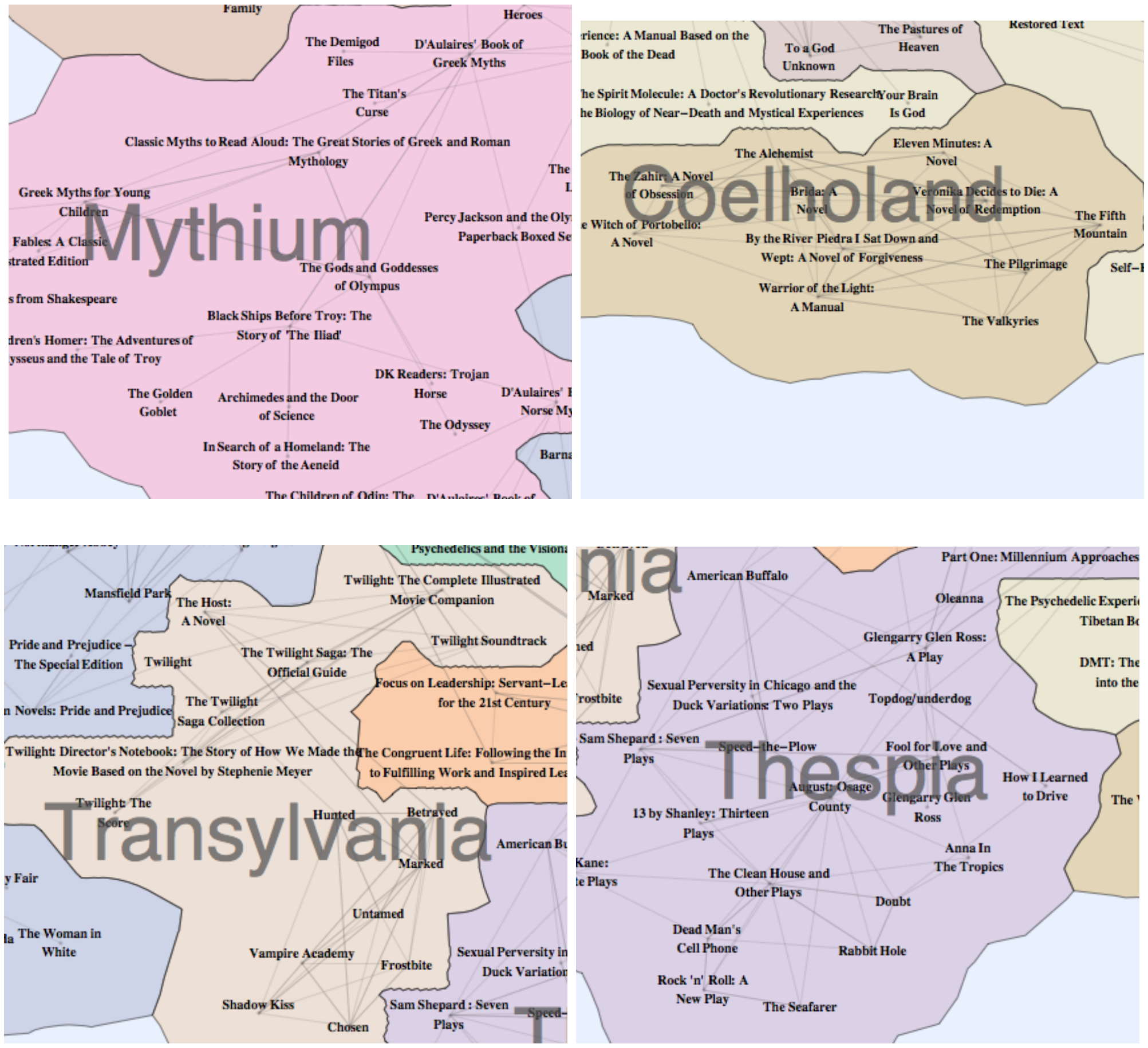

**Fig. 9.** Close-up images of various countries in BookLand'

*Help You Challenge Your Inner Critic and Celebrate Your Personal Strengths.* A related cluster of Elizabeth Gilbert books is in the same region: *Pilgrims, Last American Man* and *Stern Men.*

**KFC:** The Ken Follett Club, located to the northwest of Selfhelpistania contains some of Follett's British thrillers, *Eye of the Needle, The Man from St. Petersburg* and some of his historical fiction, *The Pillars of the Earth, Night Over Water* and *A Place Called Freedom.*

**Fringistan:** This region in the north is represented by arguably fringe work such as *Men in Black: How the Supreme Court is Destroying America, The Case Against Barack Obama: The Unlikely Rise and Unexamined Agenda of the Media's Favorite Candidate* and Bill O'Reilly's *A Bold Fresh Piece of Humanity.* Not surprisingly, the connection to the main-stream literature is through Ayn Rand's *Atlas Shrugged.*

**Feministan:** Next to Fringistan is similarly controversial cluster of books: from Naomi Wolf's *The Beauty Myth* through Ariel Levy's *Female Chauvinist Pig: Women and the Rise of Raunch Culture* to Jessica Valenti's *Full Frontal Feminism: A Young Woman's Guide to Why Feminism Matters.*

### 7.2 PotterLand

As further examples, Fig. 10 shows the continent of PotterLand, a map related to "Harry Potter and the Sorcerer's Stone". It is notable that here the clustering structure matches the layout very well, with fewer fractured countries and mostly contiguous territories. It is worth mentioning that the books and their connections in this map reflect American reading preferences (as opposed to say European or World preferences). Even the title of the the first Harry Potter book was "translated" from British English where it was *Harry Potter and the Philosopher's Stone* to American English *Harry Potter and the Sorcerer's Stone*[3] Once again, we look at some of the countries around PotterLand.

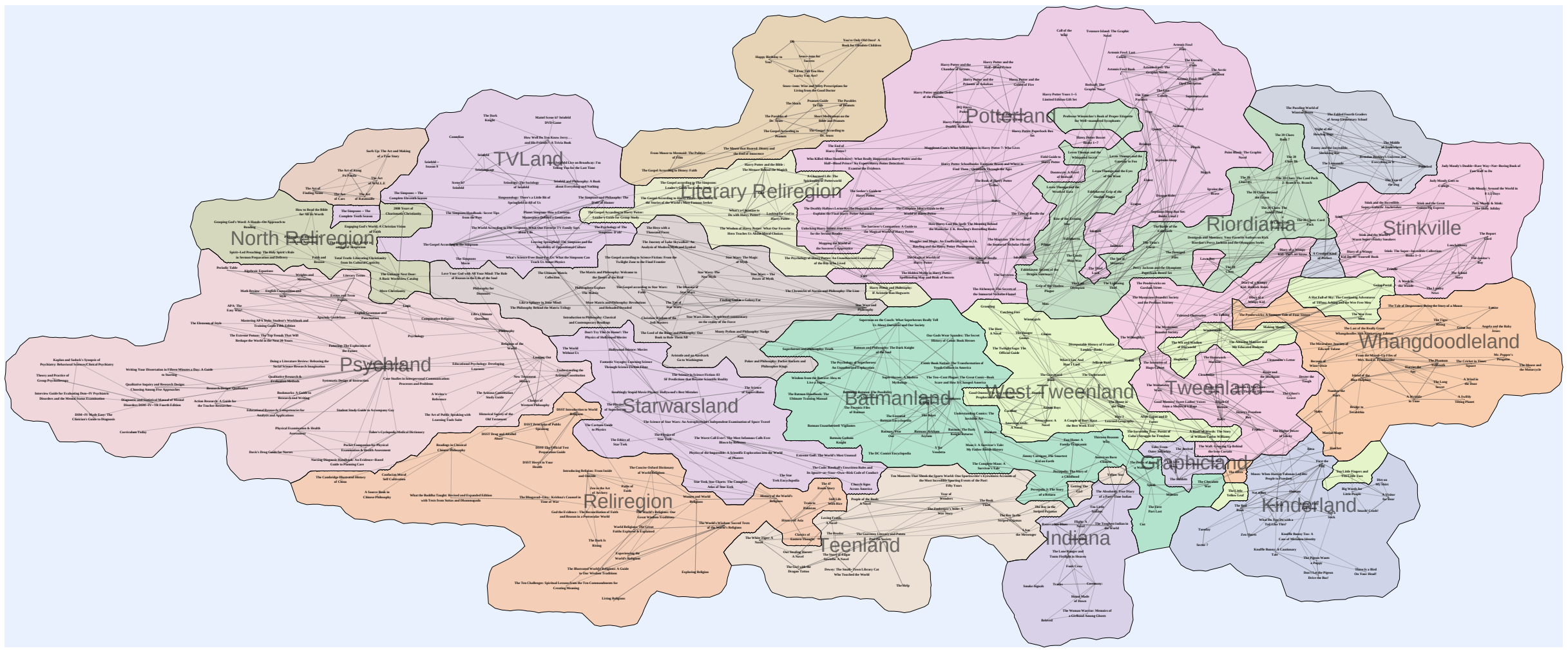

**Fig. 10.** A map of books related to "Harry Potter and the Sorcerer's Stone"

**Stinkville:** In the west is a cluster of books for 4-6 year old kids. Megan McDonald's books dominate the region with *Stink and the Great Guinea Pig Express, Stink and the Incredible Super-Galactic Jawbreaker, Stink and the World's Super-Stinky Sneakers* and the *Judy Moody* series by the same author. Andrew Clements' kids books, *Lunch Money, The Report Card* and *Frindle*, occupy the southern end of this region.

**Whangdoodleland:** Below Stinkville are books targeted at the 9-12 year old kids. Classics of the genre in this area include Katherine Paterson's *Bridge to Terabithia*, Madeline L'Engle's *A Wrinkle in Time*, Norton Juster's *Phantom Tollbooth*, E. L. Konigsburg's *From the Mixed-Up Files of Mrs. Basil E. Frankweiler*, Julie Andrews Edward's *The Last of the Really Great Whangdoodles*

**PotterLand:** The main cluster in this map is the one containing works by the British writer, J. K. Rowling. In addition to the seven books in the Harry Potter saga, there are a dozens of books about the Harry Potter books in the western part. To the east is the related cluster of the six books in the Irish writer Eoin Colfer's *Artemis Fowl* series: *Artemis Fowl, The Arctic Incident, Eternity Code, Opal Deception, Lost Colony, Time Paradox.* Directly south is the *Septimus Heap* series by the British writer Angie Sage: *Flyte, Queste, Physik, Magyk.* Finally, in the south is German writer Cornelia Funke's *Inkworld Trilogy: Inkheart, Inkspell, Inkdeath.*

**Riordiania:** Several fantasy series form the cluster to the southeast of PotterLand. The main books are Rick Riordan's *Percy Jackson and the Olympians series: The Lightning Thief, The Sea of Monsters, The Titan's Curse*, as well as a number of *39 Clues* books, another popular Riordan series. Other books in this cluster include Obert Skye's *Leven Thumps* series and Brandon Mull's *Fablehaven* series.

[3] While "philosopher's stone" is an ancient concept sought after by alchemists and scientist alike, a "sorcerer's stone" is quite meaningless without the context.

**Batmanland, Starwarsland, Startrekland, Graphicland:** Directly south of PotterLand is a cluster of Batman-related books. To the southwest is a large cluster of Star Wars books and books about Star Wars. Further south is a smaller but similar cluster of Star Trek books. Nearby is a cluster of Saun Tan's illustrated tales *Tales From Outer Suburbia, The Arrival, The Lost Thing, etc.*

**Tweenland and West Tweenland:** There are a couple of diverse clusters with contemporary books aimed at pre-teenagers. Brian Selznik's *The Invention of Hugo Cabret*, Trenton Lee Stewart's *The Mysterious Benedict Society*, and Cynthia Lord's Rules anchor Tweenland. Next door is West Tweenland with Neil Gayman's *Graveyard Book, Anansi Boys, Coraline* and books about vampires *Twilight Saga, The Host, The Hunger Games.*

**Teenland:** On the southern coast is a cluster of more mainstream books which appeal to teenagers and Oprah's Book Club. Typical examples are *White Tiger, The Guernsey Literary and Potato Peel Pie Society, People of the Book, The Story of Edgar Sawtelle, The Book Thief.*

**Indiana:** This is a cluster of books by Sherman Alexie, focusing on Native American topics *The Lone Ranger and Tonto Fistfight in Heaven, Indian Killer, The Absolutely True Diary of a Part-Time Indian.*

**Kinderland:** On the southeast coast is a collection of books targeted at the kindergarten audience: *Kitten's First Full Moon, The Pigeon Wants a Puppy* and the *Knuffle Bunny* series.

**PsychLand:** The west contains a large and diverse cluster of mostly psychological texts, anchored by *Publication Manual of the American Psychological Association.*

**Reliregion, North Reliregion, and Literary Reliregion:** the southwest coast is populated with books dealing with religion, from Christianity (*The Ten Challenges: Spiritual Lessons from the Ten Commandments for Creating Meaning*) to Buddhism (*What the Buddha Taught: Revised and Explained Edition with Texts from Suttas and Dhammapada*). In the northwest is the smaller North Reliregion cluster with books about Christianity. Immediately adjacent to PotterLand is the Literary Reliregion, dealing with religion in literature: *Looking for God in Harry Potter, Harry Potter and the Bible: The Menace Behind the Magic, What's a Christian to Do with Harry Potter.* Next to it is a related cluster of several themed books by Mark Pinsky: *The Gospel According to the Simpsons, The Gospel According to Disney*, and similarly titled books by other authors, *The Gospel According to Peanuts, The Gospel According to Dr. Seuss, The Gospel According to Harry Potter.*

**TVland:** Surrounded by religious-themed areas in the north lies a cluster of books about the popular TV shows the Simpsons and Seinfeld.

### 7.3 GDLand

Fig. 11 shows a map centered at *Graph Drawing: Algorithms for the Visualization of Graphs.* Here each vertex has a distance of 6 or less to that book. It is interesting to see how the subject quickly goes outside Mathematics and Computer Science. For example, the orange cluster in the far left contains books largely unrelated to Mathematics or Computer Science, but connected to such books via books on Game Theory.

### 7.4 GD Collaboration 1994-2008

Fig. 3 presented a map based on GD author collaboration up to 2004. More recent data from `http://gdea.informatik.uni-koeln.de/`, kindly provided to us by Michael Jünger, gives author collaboration information extending to 2008. This graph has 670 vertices and 1517 edges with a largest component of 464 nodes and 1313 edges. A map of the eight largest components in the graph is shown in Fig. 12.

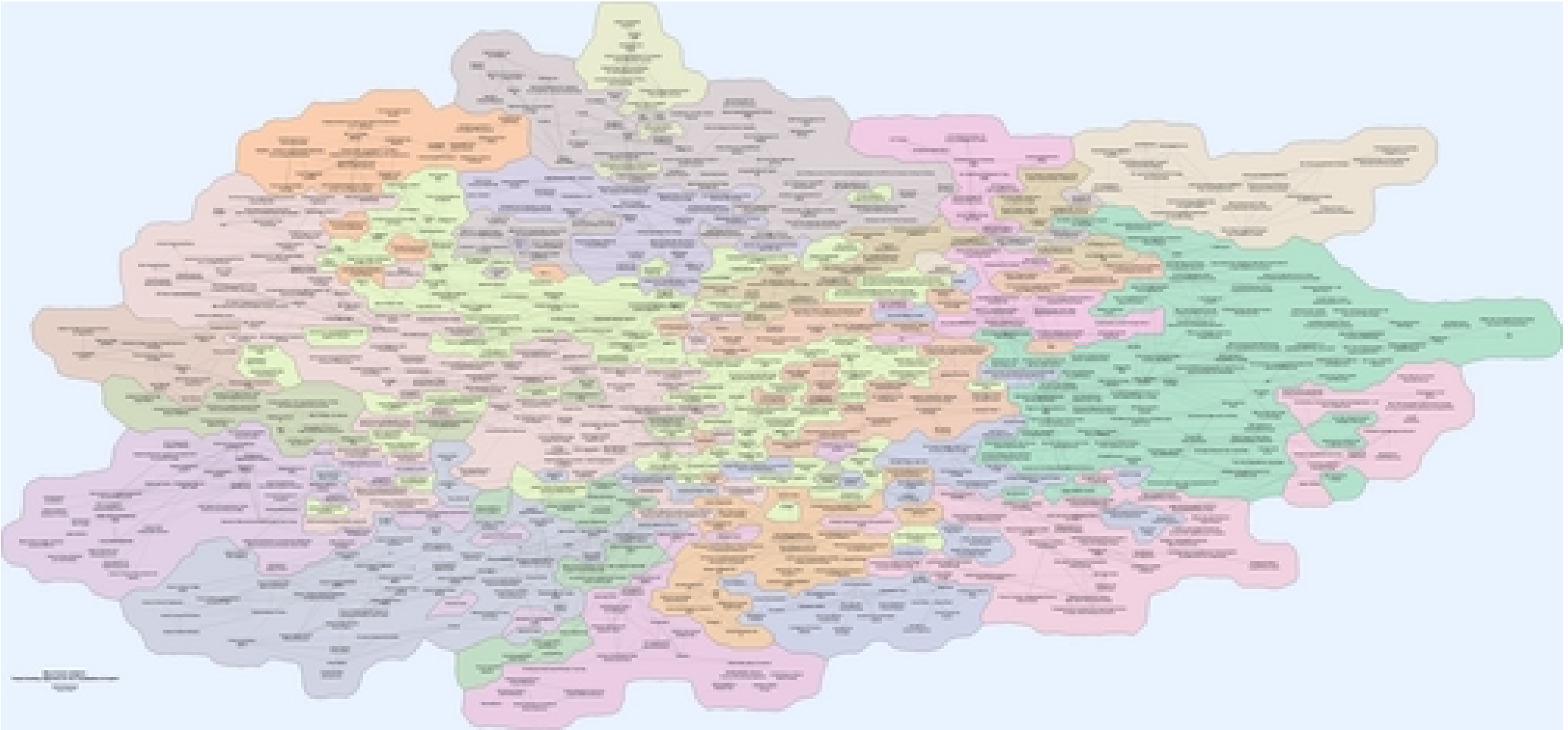

**Fig. 11.** A map of books related to *Graph Drawing: Algorithms for the Visualization of Graphs*

Due to the different origins of the two data sets it is difficult to perfectly match authors in the two graphs, and therefore provide a stable mental map. Nevertheless, when comparing these two maps of the GD community, we can observe several interesting changes. The total number of GD auhtors has grown from 508 in 2004 to 670 in 2008, or about 32%. The largest connected component has grown from 54% of all the authors in 2004 to 68% of all the authors in 2008. One of the large islands populated by Czech authors in 2004 has become part of the mainland in 2008. Similarly, the large island populated by Japanese authors has also joined the mainland, creating the peninsula on the east coast.

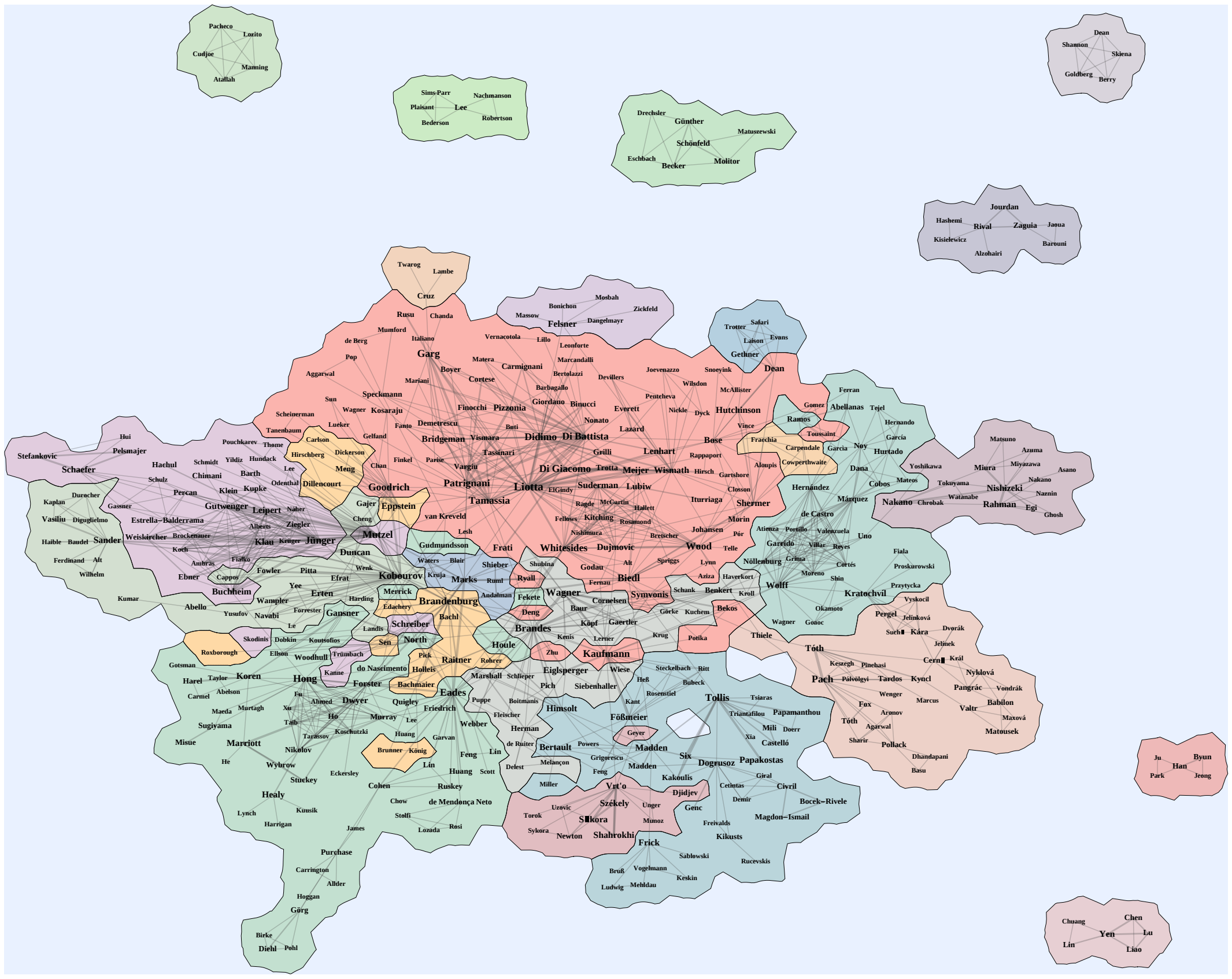

**Fig. 12.** Author collaboration map for the GD conference, 1994-2008.